\documentclass[%
reprint,
showpacs,preprintnumbers,
nofootinbib,
amsmath,amssymb,
aps,
prl,
floatfix,
]{revtex4-1}

\usepackage{graphicx}
\usepackage{dcolumn}
\usepackage{bm}

\usepackage{url}

\newcommand{\boss}[2]{\ensuremath{\rlap{\kern-2.5pt\ensuremath{\overset{\scriptscriptstyle(-)}{\phantom{#1}}}}{\ensuremath{{#1}_{#2}}}}}

\begin{document}

\preprint{\begin{tabular}{l}
\texttt{arXiv:1111.5211 [hep-ph]}
\end{tabular}}

\title{First Double-Chooz Results and the Reactor Antineutrino Anomaly}

\author{Carlo Giunti}
\email{giunti@to.infn.it}
\altaffiliation[also at ]{Department of Theoretical Physics, University of Torino, Italy}
\affiliation{INFN, Sezione di Torino, Via P. Giuria 1, I--10125 Torino, Italy}

\author{Marco Laveder}
\email{laveder@pd.infn.it}
\affiliation{Dipartimento di Fisica ``G. Galilei'', Universit\`a di Padova,
and
INFN, Sezione di Padova,
Via F. Marzolo 8, I--35131 Padova, Italy}

\date{\today}

\begin{abstract}
We investigate the possible effects of
short-baseline $\bar\nu_{e}$ disappearance implied by the reactor antineutrino anomaly
on the Double-Chooz determination of $\vartheta_{13}$
through the normalization of the initial antineutrino flux with the Bugey-4 measurement.
We show that the effects are negligible and the value of $\vartheta_{13}$ obtained by the Double-Chooz collaboration
is accurate only if $\Delta{m}^2_{41} \gtrsim 3 \, \text{eV}^2$.
For smaller values of $\Delta{m}^2_{41}$
the short-baseline oscillations are not fully averaged at Bugey-4
and
the uncertainties due to the reactor antineutrino anomaly can be of the same order of magnitude of the
intrinsic Double-Chooz uncertainties.
\end{abstract}


\maketitle

The first results \cite{DeKerret-LowNu11} of the Double-Chooz experiment
\cite{hep-ex/0606025}
led to the following result for the amplitude of
long-baseline $\bar\nu_{e}$ disappearance:
\begin{equation}
\sin^2 2 \vartheta_{13}^{\text{DC}} = 0.085 \pm 0.029 \pm 0.042
\,.
\label{st2-dch}
\end{equation}
This amplitude enters in the effective long-baseline (LBL) survival probability of $\bar\nu_{e}$
in the case of three-neutrino mixing
(see Ref.~\cite{Giunti-Kim-2007}):
\begin{equation}
P_{\bar\nu_{e}\to\bar\nu_{e}}^{\text{DC}}
=
1 - \sin^2 2 \vartheta_{13}^{\text{DC}}
\sin^2\left( \frac{\Delta{m}^2_{31}L}{4E} \right)
\,,
\label{pee-3nu}
\end{equation}
which has been assumed in the analysis of the data by the Double-Chooz collaboration
\cite{DeKerret-LowNu11}.
Here we adopt the standard parameterization of the mixing matrix,
with $|U_{e3}|=\sin\vartheta_{13}$.

An essential ingredient in the extraction of the value of
$\sin^2 2 \vartheta_{13}^{\text{DC}}$
from the data
is the normalization of the initial flux prediction on the value measured by the
Bugey-4 experiment \cite{Declais:1994ma},
since the first results of the Double-Chooz experiment
have been obtained with the far detector only \cite{DeKerret-LowNu11}.
This normalization is important because the recent recalculations of the reactor $\bar\nu_{e}$ flux
\cite{1101.2663,1106.0687}
indicate a value which is larger than that measured by
Bugey-4
and other short-baseline reactor antineutrino experiments,
leading to the reactor antineutrino anomaly \cite{1101.2755}.
The ratio of observed and theoretically predicted $\bar\nu_{e}$ flux for the Bugey-4 experiment
is \cite{1101.2755}
\begin{equation}
\frac{\phi^{\text{obs}}_{\text{Bugey-4}}}{\phi^{\text{the}}_{\text{Bugey-4}}}
=
0.942 \pm 0.042
\,,
\label{rat-bu4}
\end{equation}
and the average ratio of observed and theoretically predicted $\bar\nu_{e}$ fluxes in
short-baseline reactor antineutrino experiments
is \cite{1107.1452}
\begin{equation}
\frac{\phi^{\text{obs}}_{\text{SBL}}}{\phi^{\text{the}}_{\text{SBL}}}
=
0.946 \pm 0.024
\,,
\label{rat-sbl}
\end{equation}
which is a $2.2\sigma$ effect.
Several experiments
which could check the reactor antineutrino anomaly
have been proposed and some are already under preparation
\cite{LASSERRE-NEUTEL2011,1105.1326,1107.2335,1107.3512,1107.4766,1109.6036,1110.2983,Bowden-LowNu11,Egorov-LowNu11,YDKim-LowNu11}.

In this letter we investigate if the
short-baseline $\bar\nu_{e}$ disappearance implied by the reactor antineutrino anomaly
has an effect in the determination of $\vartheta_{13}$,
in spite of the normalization of the initial antineutrino flux with the Bugey-4 measurement.

In general,
the $\bar\nu_{e}$ flux measured in the Double-Chooz far detector is given by
\begin{equation}
\phi_{\bar\nu_{e}}^{\text{LBL}}
=
\phi_{\bar\nu_{e}}^{0}
\,
P_{\bar\nu_{e}\to\bar\nu_{e}}^{\text{LBL}}
\,,
\label{lbl-flux}
\end{equation}
where $\phi_{\bar\nu_{e}}^{0}$ is the $\bar\nu_{e}$ flux produced by the reactor
and
$P_{\bar\nu_{e}\to\bar\nu_{e}}^{\text{LBL}}$
is the effective long-baseline (LBL) $\bar\nu_{e}$ survival probability.

In the simplest framework of
3+1 neutrino mixing,
which can accommodate short-baseline neutrino oscillations together with
the well-established atmospheric (long-baseline) and solar neutrino oscillations
(see the recent Refs.~\cite{1103.4570,1107.1452,1109.4033,1110.3735,1111.1069,1111.4225} and references therein),
the effective long-baseline $\bar\nu_{e}$ survival probability is given by
\cite{0809.5076}
\begin{align}
P_{\bar\nu_{e}\to\bar\nu_{e}}^{\text{LBL}}
=
\null & \null
1 -
\cos^4 \vartheta_{14} \sin^2 2 \vartheta_{13}
\sin^2\left( \frac{\Delta{m}^2_{31}L}{4E} \right)
\nonumber
\\
\null & \null
\phantom{1} -
\frac{1}{2}
\sin^2 2 \vartheta_{14}
\,,
\label{pee-4nu}
\end{align}
where the oscillations due to
$\Delta{m}^2_{41}\gg\Delta{m}^2_{31}$
have been averaged
and we adopted a parameterization of the four-neutrino mixing matrix
in which $|U_{e3}|=\sin\vartheta_{13}\cos\vartheta_{14}$ and $|U_{e4}|=\sin\vartheta_{14}$.

\begin{figure}[t!]
\begin{center}
\includegraphics*[width=0.9\linewidth]{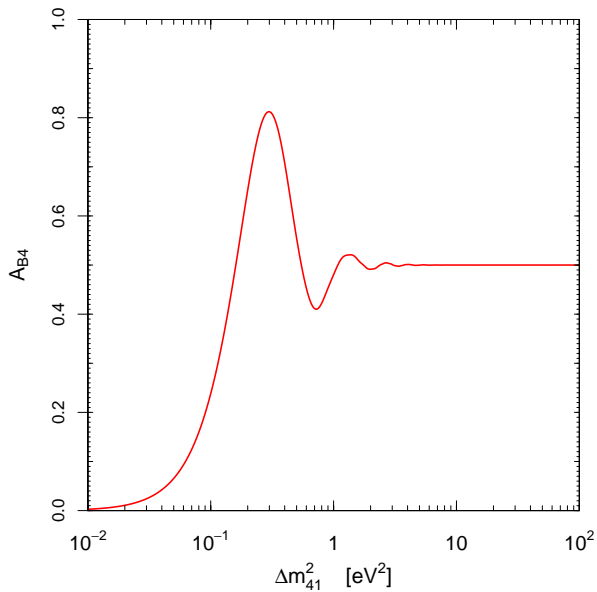}
\end{center}
\caption{ \label{ave-bu4}
Averaged value of
$
\sin^2\left( \Delta{m}^2_{41}L / 4E \right)
$
in the Bugey-4 experiment as a function of $\Delta{m}^2_{41}$.
}
\end{figure}

We can calculate the value of $\sin^2 2\vartheta_{13}$
taking into account the reactor antineutrino anomaly by noting that
in the analysis of the Double-Chooz collaboration
the $\bar\nu_{e}$ flux measured in the far detector has been fitted with
\begin{equation}
\phi_{\bar\nu_{e}}^{\text{LBL}}
=
\phi_{\bar\nu_{e}}^{\text{SBL}}
\,
P_{\bar\nu_{e}\to\bar\nu_{e}}^{\text{DC}}
\,.
\label{flux-3nu}
\end{equation}
where
$\phi_{\bar\nu_{e}}^{\text{SBL}}$
is the short-baseline $\bar\nu_{e}$ flux inferred from the Bugey-4 measurement,
taking into account the differences between the Bugey and Chooz reactors.
In the framework of 3+1 neutrino mixing,
the short-baseline $\bar\nu_{e}$ flux is given by
\begin{equation}
\phi_{\bar\nu_{e}}^{\text{SBL}}
=
\phi_{\bar\nu_{e}}^{0}
\left(
1 - A_{\text{B4}} \sin^2 2 \vartheta_{14}
\right)
\,,
\label{flux-sbl}
\end{equation}
where
$A_{\text{B4}}$
is the average of
$
\sin^2\left( \Delta{m}^2_{41}L / 4E \right)
$
in the Bugey-4 experiment.
The value of this quantity is plotted in Fig.~\ref{ave-bu4}
as a function of $\Delta{m}^2_{41}$.
One can see that
$
\sin^2\left( \Delta{m}^2_{41}L / 4E \right)
$
is fully averaged
($A_{\text{B4}}\simeq1/2$)
for $\Delta{m}^2_{41} \gtrsim 3 \, \text{eV}^2$,
but for smaller values of $\Delta{m}^2_{41}$
the average $A_{\text{B4}}$
can be significantly different from 1/2.

\begin{figure}[t!]
\begin{center}
\includegraphics*[width=0.9\linewidth]{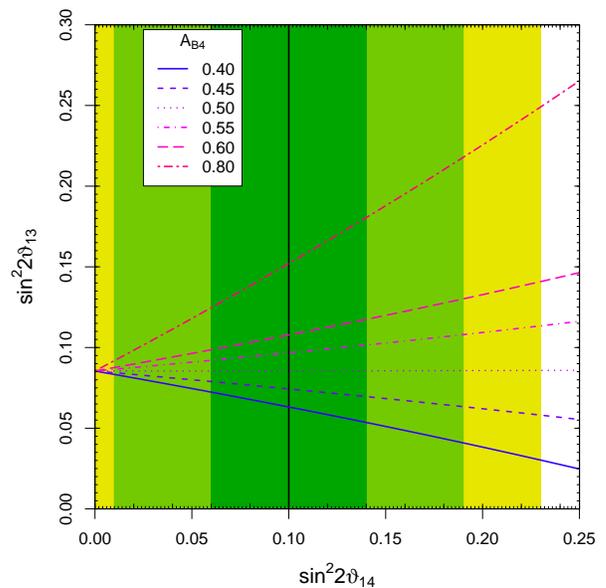}
\end{center}
\caption{ \label{plt-ab4}
$\sin^2 2 \vartheta_{13}$ as a function of $\sin^2 2 \vartheta_{14}$
obtained from Eq.~(\ref{theta}) and the best-fit value of the Double-Chooz measure in Eq.~(\ref{st2-dch})
for different values of
$A_{\text{B4}}$.
The vertical solid black line gives the best-fit value of $\sin^2 2 \vartheta_{14}$ obtained from the fit of
short-baseline reactor antineutrino data
\protect\cite{1111.1069}.
The colored vertical bands show the corresponding $1\sigma$, $2\sigma$, $3\sigma$ allowed ranges.
}
\end{figure}

Let us first consider the case of
$\Delta{m}^2_{41} \gtrsim 3 \, \text{eV}^2$,
for which
$A_{\text{B4}}\simeq1/2$.
In this case,
from Eqs.~(\ref{lbl-flux}), (\ref{pee-4nu}) and (\ref{flux-sbl}),
the $\bar\nu_{e}$ flux measured in the Double-Chooz far detector can be written as
\begin{equation}
\phi_{\bar\nu_{e}}^{\text{LBL}}
=
\phi_{\bar\nu_{e}}^{\text{SBL}}
\left[
1
-
\frac{\cos^4 \vartheta_{14} \sin^2 2 \vartheta_{13}}{1 - \frac{1}{2} \sin^2 2 \vartheta_{14}}
\sin^2\left( \frac{\Delta{m}^2_{31}L}{4E} \right)
\right]
\,.
\label{flux-4nu}
\end{equation}
Comparing Eq.~(\ref{flux-3nu}) with $P_{\bar\nu_{e}\to\bar\nu_{e}}^{\text{DC}}$
given by Eq.~(\ref{pee-3nu})
and Eq.~(\ref{flux-4nu}),
we obtain
\begin{equation}
\sin^2 2 \vartheta_{13}
=
\sin^2 2 \vartheta_{13}^{\text{DC}}
\,
\frac{1 - \frac{1}{2} \sin^2 2 \vartheta_{14}}{\cos^4 \vartheta_{14} }
\,,
\label{theta}
\end{equation}
which gives the connection between
$\vartheta_{13}^{\text{DC}}$
and
the pair $\vartheta_{13}$, $\vartheta_{14}$
for $\Delta{m}^2_{41} \gtrsim 3 \, \text{eV}^2$.

Equation~(\ref{theta}) shows that in principle
normalizing the initial neutrino flux at a value measured by a short-baseline experiment as the Bugey-4 experiment
is not sufficient to take into account the effects of short-baseline oscillations.
However, in practice the correction is small,
because the reactor antineutrino anomaly implies that $\vartheta_{14}$ is small,
which leads to
\begin{equation}
\left(1 - {\textstyle\frac{1}{2}} \sin^2 2 \vartheta_{14} \right) / \cos^4 \vartheta_{14}
=
1 + \text{O}(\vartheta_{14}^4)
\,.
\label{negligible}
\end{equation}
In this case,
the long-baseline Double-Chooz data determine the value of
$\sin^2 2 \vartheta_{13}$
independently from the value of
$\sin^2 2 \vartheta_{14}$,
which is determined by the short-baseline reactor antineutrino anomaly.

Let us now consider values of
$A_{\text{B4}}$
different from 1/2,
which can be realized if $\Delta{m}^2_{41} \lesssim 3 \, \text{eV}^2$,
as one can see from Fig.~\ref{ave-bu4}.
In this case the contribution of
$\phi_{\bar\nu_{e}}^{\text{SBL}}$
to
$\phi_{\bar\nu_{e}}^{\text{LBL}}$
cannot be factorized as in Eq.~(\ref{flux-4nu}).
Therefore, in order to find the value of
$\sin^2 2 \vartheta_{13}$
given by Double-Chooz data, we must fit these data.
We performed an approximate fit extracting the necessary information from the figures in Ref.~\cite{DeKerret-LowNu11}.
The results are shown in Fig.~\ref{plt-ab4},
where
we plotted the value of $\sin^2 2 \vartheta_{13}$ as a function of $\sin^2 2 \vartheta_{14}$
for different values of
$A_{\text{B4}}$.
Figure~\ref{plt-ab4} also shows the best fit value of $\sin^2 2 \vartheta_{14}$
and its $1\sigma$, $2\sigma$ and $3\sigma$ allowed ranges
obtained from the fit of
short-baseline reactor antineutrino data
\cite{1111.1069}.
One can see that the deviation of
$\sin^2 2 \vartheta_{13}$
from
$\sin^2 2 \vartheta_{13}^{\text{DC}}$
in the allowed band of $\sin^2 2 \vartheta_{14}$
is negligible for $A_{\text{B4}}=1/2$,
according to the discussion above.
On the other hand,
the deviation of
$\sin^2 2 \vartheta_{13}$
from
$\sin^2 2 \vartheta_{13}^{\text{DC}}$
can be relatively large for values of $A_{\text{B4}}$ different from $1/2$.
Therefore,
the uncertainty due to
short-baseline oscillations must be taken into account in the extraction of $\sin^2 2 \vartheta_{13}$
from the Double-Chooz data if $\Delta{m}^2_{41} \lesssim 3 \, \text{eV}^2$.

\begin{figure}[t!]
\begin{center}
\includegraphics*[width=0.9\linewidth]{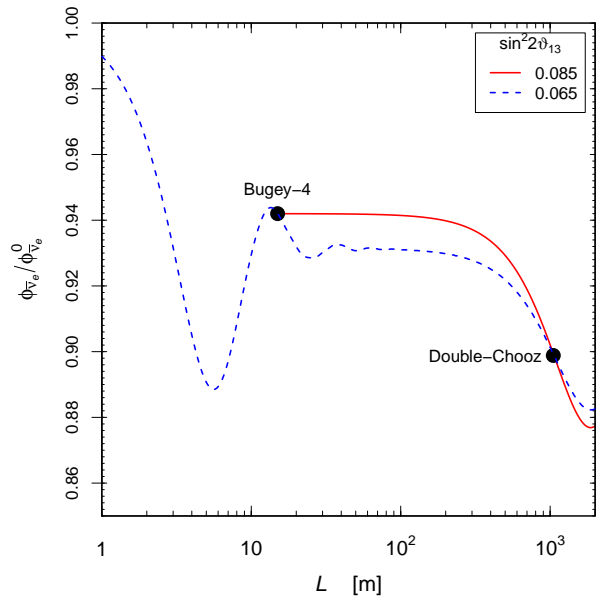}
\end{center}
\caption{ \label{len-ave-plt}
Relative suppression of the averaged reactor electron antineutrino flux as a function of distance.
The solid red curve is calculated using the three-neutrino mixing survival probability in Eq.~(\ref{pee-3nu})
with
$\Delta{m}^2_{31}=2.4\times10^{-3}\,\text{eV}^2$
and the Double-Chooz best-fit value of $\sin^2 2 \vartheta_{13}^{\text{DC}}$ in Eq.~(\ref{st2-dch})
and by normalizing the flux at the value measured by the Bugey-4 experiment at $L=15\,\text{m}$.
The blue dashed curve is calculated with the four-neutrino mixing survival probability in Eq.~(\ref{pee-4nu})
with
$\Delta{m}^2_{31}=2.4\times10^{-3}\,\text{eV}^2$,
$\sin^2 2 \vartheta_{13}=0.065$,
$\Delta{m}^2_{41}=0.8\,\text{eV}^2$ and
$\sin^2 2 \vartheta_{14}=0.14$.
The values of $\sin^2 2 \vartheta_{13}$ and $\sin^2 2 \vartheta_{14}$
have been chosen in order to fit both the Bugey-4 and Double-Chooz data points.
}
\end{figure}

\begin{figure}[t!]
\begin{center}
\includegraphics*[width=0.9\linewidth]{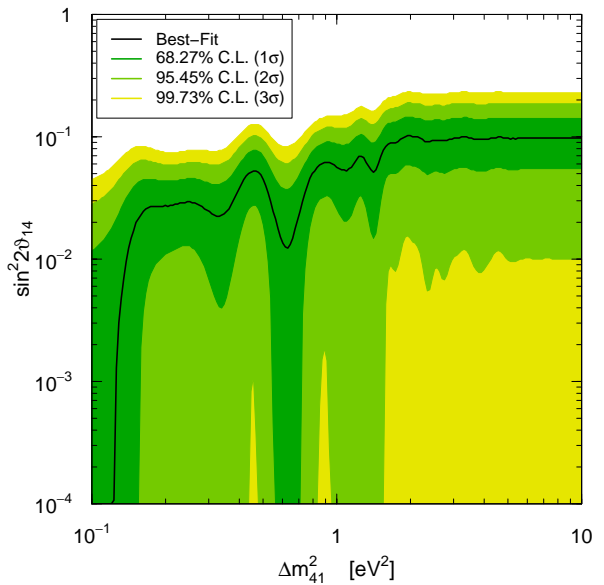}
\end{center}
\caption{ \label{plt-s14}
$\sin^2 2 \vartheta_{14}$ as a function of $\Delta{m}^2_{41}$
obtained from the fit of
short-baseline reactor antineutrino data
(see Ref.~\protect\cite{1111.1069}).
}
\end{figure}

\begin{figure}[t!]
\begin{center}
\includegraphics*[width=0.9\linewidth]{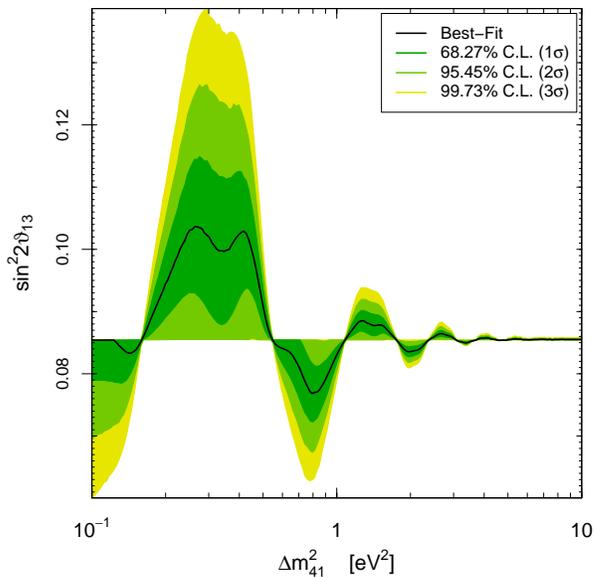}
\end{center}
\caption{ \label{plt-s13}
$\sin^2 2 \vartheta_{13}$ as a function of $\Delta{m}^2_{41}$
obtained from the best-fit value of the Double-Chooz measure in Eq.~(\ref{st2-dch})
and
$\sin^2 2 \vartheta_{14}$ in Fig.~\protect\ref{plt-s14}.
}
\end{figure}

The cause of the deviation of
$\sin^2 2 \vartheta_{13}$
from
$\sin^2 2 \vartheta_{13}^{\text{DC}}$
when the short-baseline oscillations are not fully averaged at Bugey-4
is illustrated in Fig.~\ref{len-ave-plt},
where we plotted the relative decrease of the averaged reactor electron antineutrino flux as a function of distance.
The solid red curve shows the suppression of the Bugey-4 flux
for larger distances calculated using the three-neutrino mixing survival probability in Eq.~(\ref{pee-3nu})
with the Double-Chooz best-fit value of $\sin^2 2 \vartheta_{13}^{\text{DC}}$ in Eq.~(\ref{st2-dch}).
The blue dashed curve is calculated with the four-neutrino mixing survival probability in Eq.~(\ref{pee-4nu})
and oscillation parameters chosen in order to fit both the Bugey-4 and Double-Chooz data points.
One can see that since the short-baseline oscillations are not fully averaged at Bugey-4,
the residual short-baseline oscillations at larger distances
contribute to the suppression of the flux
and the fit of the Double-Chooz data point requires a value of
$\sin^2 2 \vartheta_{13}$
which is smaller than
$\sin^2 2 \vartheta_{13}^{\text{DC}}$.

In the following we estimate the uncertainty of the determination of
$\sin^2 2 \vartheta_{13}$
from Double-Chooz data
implied by
the fit of short-baseline reactor antineutrino data
\cite{1111.1069}.

Figure~\ref{plt-s14} shows the value of
$\sin^2 2 \vartheta_{14}$
as a function of $\Delta{m}^2_{41}$
obtained from the fit of
short-baseline reactor antineutrino data
(see Ref.~\cite{1111.1069}).
Note that Fig.~\ref{plt-s14} is obtained from a one-dimensional $\chi^2$ analysis of
short-baseline reactor antineutrino data
for each fixed value of $\Delta{m}^2_{41}$.
Hence, it is different from the usual two-dimensional $\chi^2$ analyses which give allowed regions
in the
$\sin^2 2 \vartheta_{14}$--$\Delta{m}^2_{41}$
plane
(as for example Fig.~1 of Ref.~\cite{1111.1069}).
The rapid decrease of the best-fit value of
$\sin^2 2 \vartheta_{14}$
for
$\Delta{m}^2_{41} \lesssim 0.2 \, \text{eV}^2$
reflects the fact that the oscillation explanation of the reactor antineutrino anomaly
requires larger values of $\Delta{m}^2_{41}$.

Figure~\ref{plt-s13} shows the value of
$\sin^2 2 \vartheta_{13}$
as a function of $\Delta{m}^2_{41}$
obtained from our approximate fit of Double-Chooz data
and
$\sin^2 2 \vartheta_{14}$ in Fig.~\protect\ref{plt-s14}.
One can see that the deviation from
$\sin^2 2 \vartheta_{13}^{\text{DC}}$,
is smaller than about 1\%
for
$\Delta{m}^2_{41} \gtrsim 3 \, \text{eV}^2$,
in agreement with the discussion above.
On the other hand,
the deviation of
$\sin^2 2 \vartheta_{13}$
from
$\sin^2 2 \vartheta_{13}^{\text{DC}}$
can be relatively large for smaller values of $\Delta{m}^2_{41}$,
reaching about 40\% at $2\sigma$
for
$\Delta{m}^2_{41} \simeq 0.3-0.4 \, \text{eV}^2$.

Finally,
using the constraints on
$\Delta{m}^2_{41}$
and
$\sin^2 2 \vartheta_{14}$
obtained from a two-dimensional $\chi^2$ analysis of short-baseline reactor antineutrino data
\cite{1111.1069},
for the best-fit value of
$\sin^2 2 \vartheta_{13}$
in our approximate fit of Double-Chooz data
we obtain
\begin{equation}
\sin^2 2 \vartheta_{13}
=
0.084
{}^{+0.025}_{-0.010}
\,.
\label{glo}
\end{equation}
Here the uncertainties are only those due to
the analysis of short-baseline reactor antineutrino data.
Hence,
they must be added to the intrinsic Double-Chooz uncertainties in Eq.~(\ref{st2-dch}).

The result in Eq.~(\ref{glo}) shows that
the uncertainties on the determination of
$\sin^2 2 \vartheta_{13}$
due to the reactor antineutrino anomaly are comparable with the
intrinsic Double-Chooz uncertainties in Eq.~(\ref{st2-dch}).
Therefore,
the reactor antineutrino anomaly must be taken into account in the extraction of the value of
$\sin^2 2 \vartheta_{13}$
from Double-Chooz data.

In conclusion,
we have shown that if the short-baseline oscillations indicated by the reactor antineutrino anomaly exists,
in order to obtain the value of $\vartheta_{13}$
in long-baseline reactor neutrino oscillation experiments
it is not sufficient to normalize the flux at a value measured by a short-baseline experiment,
because the short-baseline oscillations may be not fully averaged at such reference point.
In the case of the first results of the Double-Chooz experiment \cite{DeKerret-LowNu11},
the flux has been normalized at the value measured by the
Bugey-4 experiment \cite{Declais:1994ma},
for which the short-baseline oscillations are fully averaged only for
$\Delta{m}^2_{41} \gtrsim 3 \, \text{eV}^2$.
We have shown that for smaller values of
$\Delta{m}^2_{41}$
the corrections due to short-baseline oscillations must be taken into account
and that a neutrino oscillation analysis of the reactor antineutrino anomaly
indicates that these corrections may be relevant.

Let us finally note that in the long-baseline reactor experiments with a near detector
which is farther from the reactor than about 100 m
(RENO \cite{1003.1391},
Daya Bay \cite{hep-ex/0701029},
Double-Chooz \cite{hep-ex/0606025})
short-baseline oscillations are fully averaged for
$\Delta{m}^2_{41}\gtrsim0.1\,\text{eV}^2$.
Therefore,
even if the reactor antineutrino anomaly is due to short-baseline oscillations,
the value of
$\sin^2 2 \vartheta_{13}$
can be extracted accurately from the data
by comparing the near and far detection rates using the three-neutrino mixing survival probability in Eq.~(\ref{pee-3nu}),
independently of the reactor antineutrino anomaly.
However,
if $\Delta{m}^2_{41}$ is sufficiently small,
the medium-baseline $\bar\nu_{e}$ flux measured in the near detector could be smaller than that measured in the
Bugey-4 experiment and in other short-baseline reactor experiments
(see Ref.~\cite{1101.2755}).
This would be a confirmation of the reactor antineutrino anomaly.

\bibliography{bibtex/nu}

\end{document}